\documentclass[pra,twocolumn,superscriptaddress,floatfix]{revtex4-2}
\usepackage[T1]{fontenc}
\usepackage{amsmath,amssymb,amsfonts}
\usepackage{graphicx}
\usepackage{times}
\usepackage{microtype}
\usepackage{physics}
\usepackage[colorlinks=true,linkcolor=blue,citecolor=blue,urlcolor=blue,breaklinks=true]{hyperref}

\newcommand{\av}{\expval}
\hypersetup{
  pdftitle={Environment-assisted squeezing in a coherently driven non-Hermitian degenerate parametric oscillator},
  pdfauthor={Muhdin Abdo Wodedo, Berihu Teklu, Konstantin G. Zloshchastiev, Jorge P. Zubelli, and Tesfay Gebremariam Tesfahannes}
}

\begin{document}

\title{Environment-assisted squeezing in a coherently driven non-Hermitian degenerate parametric oscillator}

\author{Muhdin Abdo Wodedo}
\affiliation{Department of Applied Physics, Adama Science and Technology University, 1888, Adama, Ethiopia}

\author{Berihu Teklu}
\email{berihu.gebrehiwot@ku.ac.ae}
\affiliation{Department of Mathematics, College of Computing and Mathematical Sciences, Khalifa University of Science and Technology, Abu Dhabi 127788, United Arab Emirates}
\affiliation{Center for Cyber-Physical Systems (C2PS), Khalifa University, Abu Dhabi 127788, United Arab Emirates}

\author{Konstantin G. Zloshchastiev}
\affiliation{Institute of Systems Science, Durban University of Technology, Durban 4000, South Africa}

\author{Jorge P. Zubelli}
\affiliation{Department of Mathematics, College of Computing and Mathematical Sciences, Khalifa University of Science and Technology, Abu Dhabi 127788, United Arab Emirates}
\affiliation{ADIA Lab, Level 26, Al Khatem Tower, ADGM Square, Abu Dhabi 20054, United Arab Emirates}

\author{Tesfay Gebremariam Tesfahannes}
\affiliation{Department of Physics, Arba Minch University, 21, Arba Minch, Ethiopia}

\date{July 26, 2026}

\begin{abstract}

Environment-assisted approaches to nonclassical light offer a practical path to strong squeezing in imperfect, lossy platforms. In this paper, we study a degenerate parametric oscillator in which a coherently driven cavity is coupled to a broadband squeezed reservoir via a single-port mirror. At the same time, the intracavity dynamics include non-Hermitian (gain--loss--imbalanced) terms. Within input--output theory, we obtain closed-form expressions for the steady-state quadrature variances, the output squeezing spectrum, and the power spectrum, and map their dependence on the reservoir squeeze factor, the coherent drive amplitude, and the parametric gain. We find that the non-Hermitian contributions open operating windows in which the intracavity quadrature noise is markedly suppressed below the standard quantum limit and, depending on the parameter set, either sharpen or amplify spectral squeezing and power-spectral features at the output.
The non-Hermitian coefficients are treated as effective, low-order drift parameters that describe calibrated imbalance between engineered source and sink channels after auxiliary degrees of freedom have been eliminated. The analysis is restricted to the stable Gaussian regime in which the drift matrix is stable, and the squeezed-reservoir diffusion matrix remains physical.
The results demonstrate an environment-assisted approach in which reservoir engineering and coherent driving work together to enhance squeezing. The resulting parameter maps identify experimentally testable windows, rather than a unique device prescription, for combining reservoir squeezing, coherent driving, and controlled gain/loss imbalance in cavity-QED and nonlinear photonic settings.
\end{abstract}

\maketitle

\section{Introduction}\label{sec1}

The generation of squeezed states of light has been extensively investigated both theoretically and realized experimentally \cite{b1,Purdy2013,Ma:20,Haitham2021,Colin2021,Ingrid2021,Xing2023}. A squeezed state satisfies the uncertainty relation while redistributing quantum noise between two conjugate quadratures: the variance in one quadrature is reduced below the vacuum (or coherent-state) level at the cost of increased fluctuations in the other \cite{walls1983squeezed}. Accordingly, this feature of squeezed light has potential applications in the precision of optical measurements, optical communications, the detection of weak signals, and laser interferometers \cite{kimble1987generation,walls1994generation,scully1997quantum,Vahlbruch2016,SCHNABEL20171,Aldana2014,Alessandro,Simone25}.

The degenerate optical parametric oscillator (DOPO) remains a canonical platform where quantum fluctuations can be engineered, and whose threshold behavior plays directly a role in the attainable noise reduction and spectral properties \cite{collett_gardiner_1984,gardiner_collett_1985,CavesSchumakerI1985,CavesSchumakerII1985,Wu1986,Wu1987,Andersen_2016,az92}. It has been shown that the standard 50\% limit of the second-order squeezing inside the cavity can be surpassed by replacing an ordinary vacuum reservoir with a squeezed-vacuum reservoir. Furthermore, a DOPO driven by coherent light and coupled to a squeezed-vacuum reservoir has been examined in detail \cite{TEKLU2006}. The coherent drive enhances the mean photon number of the cavity mode without degrading its squeezing properties. In addition, reservoir-engineered optomechanical cavities incorporating a nondegenerate parametric amplifier have also been investigated \cite{WODEDO2025108364,Wodedo11142662}.

Non-Hermitian (NH) Hamiltonians have numerous applications in many areas of physics, including studies of Feshbach resonances and decaying states, quantum transport and scattering by complex potentials, multi-photon ionization, free-electron lasers, and optical resonators and waveguides. However, their broadest area of application is the theory of open quantum systems, where the anti-Hermitian part arises as a result of the interaction of systems with their environment \cite{suu54,kor64,wong67,heg93,bas93,rotter09,reiter,sz13,kar14,sz14,sz14cor,sz16,z15,z16,gom17,MiriAlu2019,KawabataPRX2019,Ashida2020,BergholtzRMP2021,chw22,Luo2022,z24jmo,Du2025}. More recently, quantum-compatible NH strategies, e.g., pseudo-(anti-)\(\mathcal{PT}\) configurations have been proposed to enhance squeezing and sensing while taming Langevin noise \,\cite{Luo2022}. Despite the long history of the field, the core formalism of the non-Hermitian quantum dynamics is still a subject of active research from the viewpoint of the density-operator approach
\cite{sz13,kar14,sz14,sz14cor,sz16,z15,z24uni}. Experimental realization and applications of non-Hermitian systems in optical micro-resonators and photonic platforms have been reviewed and demonstrated \cite{peng2014,Feng2017}. Recently, the exploration of non-Hermitian quantum effects by bridging the fields of atomic physics, non-Hermitian optics, quantum information, and reservoir engineering has been investigated \cite{Cao2020}. These findings have potential applications that include novel quantum light sources, quantum information processing and sensing, and generalization to correlated many-body systems. More recently, quantum dynamics with stochastic non-Hermitian Hamiltonians have been analyzed \cite{Pablo25}. These results show that adding noise can improve control of the dynamics, with a greater diversity of steady states and the possibility of state purification.

For readers more familiar with standard quantum-optical master equations, we emphasize that the NH Hamiltonian used below is an effective description of the deterministic part of the reduced cavity dynamics. It should not be interpreted as a replacement for the Lindblad noise channels required by quantum mechanics. Instead, it represents the drift generated after auxiliary lossy or amplifying degrees of freedom, absorptive elements, or measurement-conditioned channels have been eliminated. The quantum fluctuations associated with the accessible port are still included explicitly through the input--output squeezed reservoir. In this sense, the parameters introduced below are phenomenological, measurable coefficients of the linearized drift matrix, and their admissible range is fixed by stability and by the positivity of the resulting covariance matrix.

In this paper, we show that environment-assisted NH terms provide a simple and robust control parameter to reshape both steady-state quadrature fluctuations and the output squeezing and power spectra of a DOPO\@. We demonstrate regimes in which the amplitude-quadrature variance dips below shot noise more than in the Hermitian limit, accompanied by characteristic spectral signatures at the output. These results consolidate NH control as a practical tool for upgrading quantum-limited sensing and information processing, while remaining compatible with standard input--output formalisms \,\cite{gardiner_collett_1985}.

The structure of this paper is as follows. In Sec.~\ref{sec:2}, we present the model and derive the mean (classical) dynamical equations from the system's master equation for the cavity mode and the trace of the density operator. The solutions of these equations are then used in Sec.~\ref{sec:3} to compute the quadrature fluctuations, in Sec.~\ref{sec:4} to evaluate the squeezing spectrum, and in Sec.~\ref{sec:5} to determine the power spectrum of the cavity mode. Finally, we summarize our main results in Sec.~\ref{sec:6}.

\section{Model}\label{sec:2}

\begin{figure}[tbp]
\centering
\includegraphics[width=\columnwidth]{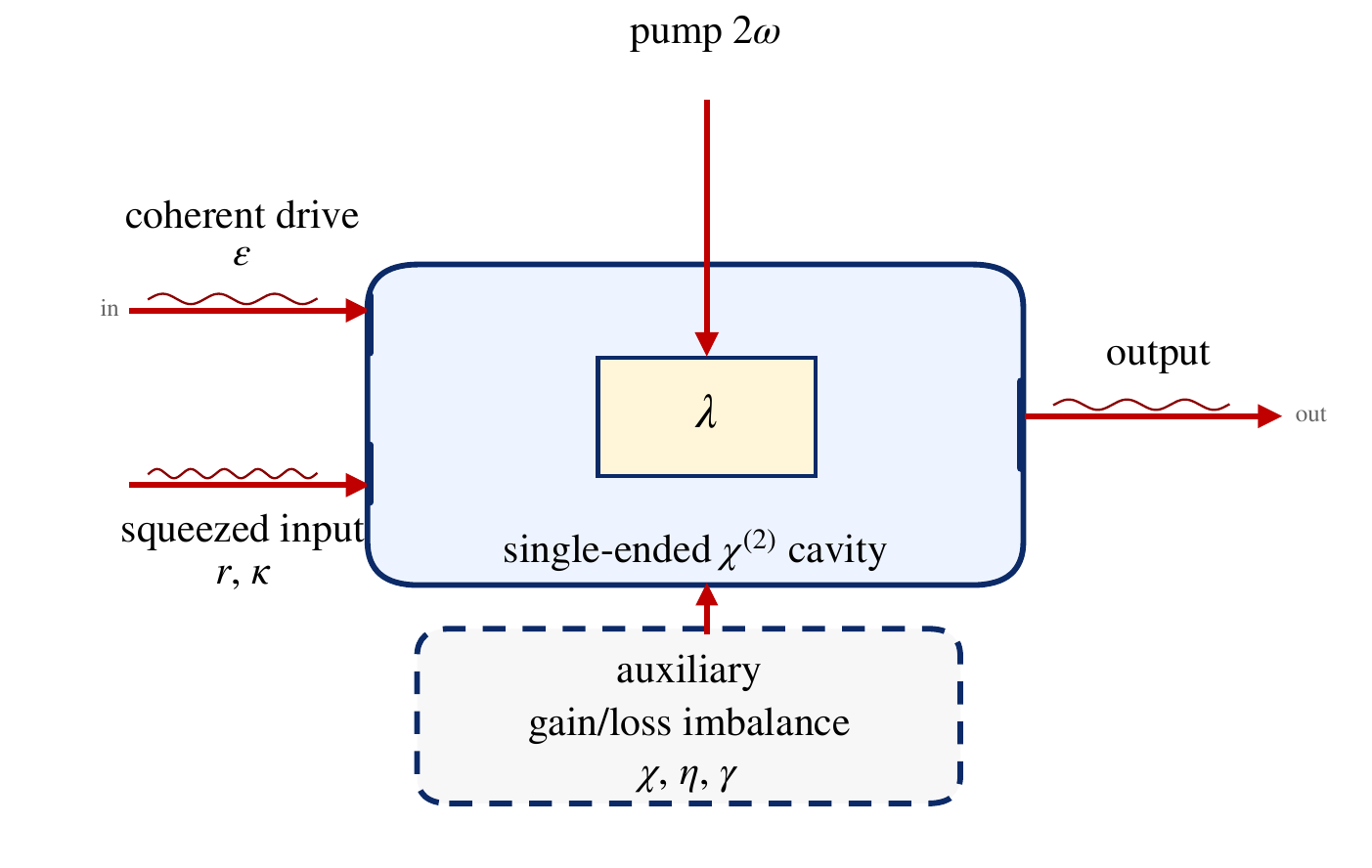}
\caption{Schematic of the effective setup considered in this work. The $\chi^{(2)}$ element provides the Hermitian parametric coupling $\lambda$, the resonant coherent input sets the drive amplitude $\varepsilon$, the squeezed-reservoir port is characterized by the squeeze parameter $r$ and decay rate $\kappa$, and a calibrated auxiliary gain/loss imbalance yields the effective non-Hermitian drift parameters $\chi$, $\eta$, and $\gamma$.}
\label{fig:setup}
\end{figure}

We consider a degenerate parametric oscillator comprising a single cavity mode driven by a coherent laser and coupled to a squeezed-vacuum reservoir, including non-Hermitian effects. Our analysis focuses on steady-state properties of both the intracavity field and the emitted radiation. In a degenerate parametric oscillator, a pump photon at frequency $\omega_p = 2\omega$ is down-converted into a pair of signal photons at frequency $\omega$.
A possible physical realization is sketched in Fig.~\ref{fig:setup}. The mode at $\omega$ is supported by a single-ended nonlinear cavity containing a $\chi^{(2)}$ medium; the pump at $2\omega$ sets the coherent parametric coupling $\lambda$, a weak resonant laser sets the drive amplitude $\varepsilon$, and the input mirror couples the cavity to a broadband squeezed vacuum characterized by $r$ and rate $\kappa$. The NH parameters are modeled as effective drift terms produced by a calibrated auxiliary gain/loss imbalance, for example, an absorptive or amplifying element phase-locked to the pump/drive, or an auxiliary mode that has been adiabatically eliminated. Their values are therefore not arbitrary: in an experiment, they would be inferred from the measured linear response or from homodyne spectra by fitting the drift matrix in Eqs.~\eqref{eq:17}--\eqref{eq:21}.
We use a non-Hermitian Hamiltonian of the form
\begin{equation}\label{eq:1}
    \hat{\cal H}=\hat H -i \hat\Gamma,
\end{equation}
where both $\hat H$ and $\hat\Gamma$ are Hermitian operators ($\hat\Gamma$ is often called the \textit{decay operator}). In the interaction picture, we write
\begin{eqnarray}\label{eq:2}
    \hat H
&=&
\frac{i \lambda}{2}
\left(
\hat a^2 - \hat a^{\dagger 2}
\right)
+
i \varepsilon
\left(
\hat a - \hat a^{\dagger}
\right)
,
\end{eqnarray}
{\begin{eqnarray}\label{eq:3}
\hat\Gamma&=&\frac{i \chi}{2}
\left(\hat a^2 - \hat a^{\dagger 2}
\right)+
\frac{\eta}{2}\left(\hat a + \hat a^{\dagger}\right) +\frac{\gamma}{2} \hat I,
\end{eqnarray}}
where
$\lambda$, $\chi$, $\varepsilon$, $\eta$, $\gamma$ are real-valued parameters, and $\hat I$ is the unit operator.
The first term in Eq.~\eqref{eq:2} describes the coherent parametric down-conversion in which a pump photon at frequency $2\omega$ is converted into two photons at frequency $\omega$, while the pump is treated classically. The second term accounts for the coherent driving of the cavity mode by an external laser.
Equation~(\ref{eq:3}) is a minimal, Gaussian-preserving decay operator. The coefficient $\chi$ is the NH counterpart of the parametric coupling and represents an imbalance between pair-creation and pair-absorption drift channels. The coefficient $\eta$ is the NH counterpart of the coherent drive and represents a linear source/sink imbalance in phase with the drive. The scalar term $\gamma/2$ changes the trace of the auxiliary, non-normalized density matrix but drops out of normalized expectation values. These terms are therefore not additional Lindblad dissipators and are not assumed to be mandatory for every open optical cavity; rather, following the density-matrix formalism for NH dynamics, they provide an effective way to parametrize a calibrated deterministic gain/loss imbalance, while the quantum noise is explicitly included in the reservoir dissipator below. The values used in the plots, $|\chi|,|\eta| \lesssim 0.1\kappa$, are chosen as weak-to-moderate perturbations of the cavity linewidth so that the linearized dynamics remain stable.

Using the non-normalized density-matrix formalism summarized in~\ref{appendix}, together with the standard broadband squeezed-reservoir dissipator of input--output theory \cite{gardiner_collett_1985,CavesSchumakerI1985,CavesSchumakerII1985,walls2008input,TEKLU2006}, we obtain the following master equation for an auxiliary (non-normalized) density matrix $\rho$:
\begin{eqnarray}\label{eq:4}
 \frac{d}{d t} \hat \rho
&=&
\frac{\Lambda^*}{2}
\left(
\hat a^2 - \hat a^{\dagger 2}
\right) \hat \rho
-
\frac{\Lambda}{2}
\hat \rho
\left(
\hat a^2 - \hat a^{\dagger 2}
\right)
-
\varepsilon_+
\left(
\hat a \hat \rho
+
\hat \rho \hat a^{\dagger}
\right)
\nonumber \\ &&+
\varepsilon_-
\left(
\hat \rho \hat a
+
\hat a^{\dagger} \hat \rho
\right) - \gamma \hat \rho +
\frac{\kappa}{2}
\hat{\cal D} ({\hat \rho}),
\end{eqnarray}
with $\Lambda = \lambda + i \chi$,
{$\varepsilon_\pm = \varepsilon \pm \eta/2$}, $\kappa$ is the cavity decay rate, while the dissipator operator is defined as
\begin{equation}
\label{eq:5}
\begin{aligned}
\hat{\mathcal{D}}(\hat{\rho})
={}& (N+1)
\Bigl(
    2\hat{a}\hat{\rho}\hat{a}^{\dagger}
    -\hat{a}^{\dagger}\hat{a}\hat{\rho}
    -\hat{\rho}\hat{a}^{\dagger}\hat{a}
\Bigr)
\\
&+N
\Bigl(
    2\hat{a}^{\dagger}\hat{\rho}\hat{a}
    -\hat{a}\hat{a}^{\dagger}\hat{\rho}
    -\hat{\rho}\hat{a}\hat{a}^{\dagger}
\Bigr)
\\
&+M
\Bigl(
    2\hat{a}\hat{\rho}\hat{a}
    +2\hat{a}^{\dagger}\hat{\rho}\hat{a}^{\dagger}\\
   & -\hat{a}^{2}\hat{\rho}
    -\hat{\rho}\hat{a}^{2}
    -(\hat{a}^{\dagger})^{2}\hat{\rho}
    -\hat{\rho}(\hat{a}^{\dagger})^{2}
\Bigr).
\end{aligned}
\end{equation}

where
\begin{equation}
\label{eq:6}
\begin{aligned}
N &= \sinh^{2}(r), \\
M &= \sinh(r)\cosh(r)
   = \frac{1}{2}\sinh(2r).
\end{aligned}
\end{equation}

Here $r$ denotes the squeeze parameter of the broadband environmental mode that is resonant with the signal mode of the cavity; the squeeze phase has been chosen as the quadrature reference phase, so that $M$ is real. For a pure squeezed vacuum, $M^2=N(N+1)$, which ensures that the reservoir correlations define a physical Gaussian input noise.

The main parameters of the model and their experimental interpretation are summarized in Table~\ref{tab:parameters}.
\begin{table*}[t]
\centering
\caption{Summary of the main parameters and their physical interpretation in the effective model.}
{\begin{tabular}{p{0.12\textwidth}p{0.31\textwidth}p{0.48\textwidth}}
\hline
Parameter & Meaning & Experimental control or calibration \\
\hline
$\lambda$ & Hermitian parametric coupling of the $\chi^{(2)}$ nonlinear crystal & Pump amplitude at $2\omega$; below threshold in the Hermitian limit, $\lambda<\kappa/2$ \\
$\varepsilon$ & Resonant coherent drive amplitude & Input laser amplitude and phase at the signal frequency $\omega$ \\
$\kappa$ & Coupling/decay rate through the input--output mirror & Mirror transmission and total linewidth used for normalization \\
$r$ & Squeeze parameter of the resonant broadband input reservoir & External squeezed-light source; $r=0$ gives ordinary vacuum \\
$\chi$ & NH parametric drift imbalance & Calibrated auxiliary loss/gain or eliminated mode that modifies the pair process \\
$\eta$ & NH coherent-drive drift imbalance & Calibrated source/sink imbalance phase-locked to the coherent drive \\
$\gamma$ & Uniform trace loss/gain of the auxiliary density operator & Affects normalization of $\rho$ but not normalized observables \\
\hline
\end{tabular}}
\label{tab:parameters}
\end{table*}

From the master equation, Eq.~\eqref{eq:4}, we obtain the mean (classical) dynamics expressions as

\begin{align}
   \frac{d}{d t} \av {\hat a}
=&-i \chi\av{\hat a^3}+i \chi\av{\hat a^{\dagger 2}\hat a}-\eta\av{\hat a^2}-\eta\av{\hat a^{\dagger}\hat a} \nonumber\\&-\frac{\kappa}{2}\av{\hat a}-\Lambda^{*}\av{\hat a^{\dagger}}-\gamma \av{\hat a}+\varepsilon_{-}{\rm Tr}(\hat\rho), \label{eq:7}
\\
    \frac{d}{d t} {\rm Tr}(\hat \rho)=& -i \chi\av{\hat a^2}+i \chi\av{\hat a^{\dagger 2}}-\eta\av{\hat a}-\eta\av{\hat a^{\dagger}} \nonumber\\&-\gamma{\rm Tr}(\hat \rho)
    ,
    \label{eq:8}
    \end{align}
and similarly for $\langle \hat a^\dagger \rangle$.

Because our system includes both coherent driving and reservoir-induced gain, the probability space must remain normalized in the physical description. Therefore, we introduce the normalized density matrix $\hat\rho'$, which will serve as the physical statistical operator from now on.

Using Eqs.~\eqref{eq:7} and \eqref{eq:8}
and the relations in Appendix~\ref{appendix},
we obtain rate equations for our primary physical observables.
For the mean field we find
\begin{align}
\frac{d}{dt}\av{\hat a}'
={}&-i\chi\left(\av{\hat a^3}'-\av{\hat a^{\dagger 2}\hat a}'\right)\nonumber\\
&-\eta\left(\av{\hat a^2}'+\av{\hat a^{\dagger}\hat a}'\right)\nonumber\\
&+i\chi\av{\hat a}'\left(\av{\hat a^2}'-\av{\hat a^{\dagger 2}}'\right)\nonumber\\
&+\eta\av{\hat a}'\left(\av{\hat a}'+\av{\hat a^{\dagger}}'\right)\nonumber\\
&-\frac{\kappa}{2}\av{\hat a}'-\Lambda^*\av{\hat a^{\dagger}}'
+\varepsilon_-,
\label{eq:9}
\end{align}
and similarly for the creation operator.
Notice that the mean value computed with respect to the normalized density matrix does not depend on $\gamma$. This is a general feature of non-Hermitian Hamiltonian evolutions with normalized density matrices: unit-matrix components of the decay operator $\hat\Gamma$ or the energy operator $\hat H$ do not contribute to the dynamics.
The spectra below are evaluated only when the linearized drift matrix has eigenvalues with negative real parts. In this stable regime, the covariance matrix is finite, and the symmetrized intracavity and output spectra are physically meaningful observables accessible by homodyne detection. If the NH coefficients were chosen so large that this stability condition failed, the linearized steady-state spectra would not be interpreted as physical steady-state squeezing spectra.

\section{Quadrature fluctuations}\label{sec:3}

We now calculate the quadrature variance and squeezing spectrum for the cavity mode under consideration. Following the standard linearization approach, we decompose the cavity
field operator into its steady-state mean value and its quantum fluctuations,
\begin{align}\label{eq:10}
    \hat{a} =\alpha+\delta\hat{a}
\end{align}
where $\alpha=\av{a}'$ denotes the classical coherent amplitude in the presence of non-Hermitian parameter $\Gamma$ and $\delta\hat{a}$ is the fluctuation operator with zero mean value. The classical dynamics of $\alpha$ and $\alpha^*$ follow directly from Eq.~\eqref{eq:9} and its Hermitian conjugate:
\begin{align}
\frac{d\alpha }{d t}
=&-\frac{\kappa}{2}\alpha-\Lambda^{*} \alpha^{*} + \varepsilon_-,\label{eq:11} \\
\frac{d\alpha^{*} }{d t}
=&-\frac{\kappa}{2}\alpha^{*}-\Lambda \alpha +\varepsilon_-,\label{eq:12}
\end{align}
Furthermore, at steady state ($t \to \infty$), one can take $\frac{d\alpha (t)}{d t}=0 $, and $\frac{d\alpha^{*} (t)}{d t}=0 $. Setting the time derivatives in Eqs.~\eqref{eq:11} and \eqref{eq:12} to zero gives
\begin{align}\label{eq:13}
\alpha=\frac{2(\kappa-2\Lambda^*)\varepsilon_-}{\kappa^2-|2\Lambda|^2 }
\end{align}
The quantum dynamics of the fluctuations $\delta\hat a$ and $\delta\hat a^{\dagger} $ can be obtained using the stochastic differential (quantum Langevin) form of Eq.~\eqref{eq:9}, yielding
\begin{align}
  \frac{d}{d t} \delta {\hat a}
=&i \chi\delta {\hat a}\big(
\alpha^{2}-\alpha^{*2}\big)
+\eta\delta {\hat a}\big(\alpha+\alpha^{*}\big)\nonumber\\&-\frac{\kappa}{2}\delta{\hat a}-(\lambda-i\chi)\delta{\hat a^{\dagger}}+ \sqrt{\kappa}\hat a_{\mathrm{in}},\label{eq:14}
\\
\frac{d}{d t} \delta\hat a^{\dagger}
=&i \chi\delta {\hat a}^{\dagger}\big(
\alpha^{2}-\alpha^{*2}\big)
+\eta\delta {\hat a}^{\dagger}(\alpha+\alpha^*)\nonumber\\&
-\frac{\kappa}{2}\delta{\hat a^{\dagger}}-(\lambda+i\chi)\delta{\hat a}+ \sqrt{\kappa}\hat a^{\dagger}_{\mathrm{in}},\label{eq:15}
\end{align}
where $\hat a_{\mathrm{in}}$ indicates the input squeezed vacuum noise operator with central frequency resonant to the cavity mode frequency. It has zero mean value and nonzero time-domain correlation functions
\begin{align}\label{eq:16}
\begin{split}
    \av{\delta a^{\dagger}_{\mathrm{in}}(t) \delta a_{\mathrm{in}}(t')} =&N\delta(t-t'),\\
 \av{\delta a_{\mathrm{in}}(t) \delta a^{\dagger}_{\mathrm{in}}(t')} =&(N+1)\delta(t-t'),\\
 \av{\delta a_{\mathrm{in}}(t) \delta a_{\mathrm{in}}(t')} =&
 \av{\delta a^{\dagger}_{\mathrm{in}}(t) \delta a^{\dagger}_{\mathrm{in}}(t')} =-M\delta(t-t'),\\
\end{split}
\end{align}

To obtain quantitative measures we introduce quadrature operators and their associated noise operators as $\delta\hat q=\frac{1}{\sqrt{2}}(\delta\hat a^{\dagger}+\delta\hat a),\delta\hat p=\frac{i}{\sqrt{2}}(\delta\hat a^{\dagger}-\delta\hat a), \hat q_{\mathrm{in}}=\frac{1}{\sqrt{2}}(\delta\hat a^{\dagger}_{\mathrm{in}}+\delta\hat a_{\mathrm{in}}), \hat p_{\mathrm{in}}=\frac{i}{\sqrt{2}}(\delta\hat a^{\dagger}_{\mathrm{in}}-\delta\hat a_{\mathrm{in}})$. In terms of these quadratures, Eqs.~\eqref{eq:14} and \eqref{eq:15} become
 \begin{align}
 \frac{d}{d t} \delta\hat q
=&-\bigg( \frac{\kappa}{2}+\lambda -i \chi\big(
\alpha^{2}-\alpha^{*2}\big)-\eta\big(\alpha+\alpha^{*}\big)\bigg)\delta \hat q\nonumber\\&+
\chi\delta\hat p+ \sqrt{\kappa}\hat q_{\mathrm{in}},\label{eq:17}\\
\frac{d}{d t} \delta\hat p
=&-\bigg( \frac{\kappa}{2}-\lambda -i \chi\big(
\alpha^{2}-\alpha^{*2}\big)-\eta\big(\alpha+\alpha^{*}\big)\bigg)\delta\hat p \nonumber\\& +\chi\delta \hat q+ \sqrt{\kappa}\hat p_{\mathrm{in}}.\label{eq:18}
\end{align}

The dynamics can be written compactly as a first-order matrix differential equation,

\begin{equation}\label{eq:19}
      \dot{\mathbf{u}}(t) =\mathbf{J} \mathbf{u}(t)+\mathbf{n}(t)
\end{equation}
where $\mathbf{u}(t)=(\delta\hat q(t),\delta\hat p(t))^{\mathsf T}$ is a column vector of quadrature fluctuations, $\mathbf{n}(t)=(\sqrt{\kappa}\hat q_{\mathrm{in}}(t),\sqrt{\kappa}\hat p_{\mathrm{in}}(t))^{\mathsf T}$ is a column vector of noise quadrature fluctuations and the drift matrix $\mathbf{J}$ containing the system information is given by
\begin{equation} \label{eq:20}
    \mathbf{J}=\begin{pmatrix}
  J_{11} & J_{12}\\
   J_{21}&J_{22} \\
\end{pmatrix}
\end{equation}
where
\begin{align}\label{eq:21}
\begin{split}
J_{11}=&  -\bigg( \frac{\kappa}{2}+\lambda -i \chi \big(\alpha^{2} - \alpha^{*2}\big) -\eta\big(\alpha+\alpha^{*}\big)\bigg)\\
J_{12}=&\chi\\
J_{21}=&\chi\\
 J_{22}=&-\bigg( \frac{\kappa}{2}-\lambda -i \chi\big(
\alpha^{2}-\alpha^{*2}\big)-\eta\big(\alpha+\alpha^{*}\big)\bigg)  \\
\end{split}
\end{align}
Taking the Fourier transform of Eq.~\eqref{eq:19}, using \(f(t)=\frac{1}{2\pi}\int_{-\infty}^{\infty}f(\omega)e^{-i\omega t}d\omega\) and \(\partial_t f(t) = -i \omega f(\omega)\), we obtain

\begin{align}\label{eq:22}
\mathbf{u}(\omega)=(-i\omega \mathbf{I}-\mathbf{J})^{-1}\mathbf{n}(\omega),
\end{align}
with $\mathbf{I}$ the $2\times 2$ identity matrix. The amplitude and phase quadrature fluctuations of the cavity field can thus be written as
\begin{align}\label{eq:23}
\begin{split}
    \delta \hat q (\omega) =& A_1(\omega) \hat q_{\mathrm{in}}(\omega) + B_1(\omega) \hat p_{\mathrm{in}}(\omega),\\
\delta \hat p(\omega) = &A_2(\omega)\hat q_{\mathrm{in}}(\omega) + B_2(\omega) \hat p_{\mathrm{in}}(\omega)
\end{split}
\end{align}
where
\begin{align}\label{eq:24}
\begin{split}
    A_1(\omega) =&-\frac{\sqrt{\kappa}}{D(\omega)}[J_{22}+i\omega],\\
    B_1(\omega)=&\frac{\sqrt{\kappa}}{D(\omega)}J_{12} ,\\
A_2(\omega)=&\frac{\sqrt{\kappa}}{D(\omega)}J_{21} ,\\
B_2(\omega)= &-\frac{\sqrt{\kappa}}{D(\omega)}[J_{11}+i\omega],\\
\end{split}
\end{align}
with $D(\omega)=(J_{11}+i\omega)(J_{22}+i\omega)-J_{12}J_{21}$.
The spectra of the quadrature fluctuations are defined as
\begin{align}
    S_{q} (\omega) =\frac{1}{2\pi}\int_{-\infty}^{\infty}d\omega'e^{i(\omega+\omega')t}\frac{1}{2}\bigg(&\av{\delta q(\omega)\delta q(\omega')}\nonumber\\&+\av{\delta q(\omega')\delta q(\omega)} \bigg), \label{eq:25} \\
   S_{p} (\omega) =\frac{1}{2\pi}\int_{-\infty}^{\infty}d\omega'e^{i(\omega+\omega')t}\frac{1}{2}\bigg(&\av{\delta p(\omega)\delta p(\omega')}\nonumber\\&+\av{\delta p(\omega')\delta p(\omega)} \bigg).\label{eq:26}
\end{align}

Using the Fourier-domain noise correlations derived from Eq.~\eqref{eq:16},
\begin{align}\label{eq:27}
\begin{split}
    \av{\delta a^{\dagger}_{\mathrm{in}}(\omega) \delta a_{\mathrm{in}}(\omega')} =&2\pi N\delta(\omega+\omega'),\\
 \av{\delta a_{\mathrm{in}}(\omega) \delta a^{\dagger}_{\mathrm{in}}(\omega')} =&2\pi (N+1)\delta(\omega+\omega'),\\
 \av{\delta a_{\mathrm{in}}(\omega) \delta a_{\mathrm{in}}(\omega')} =&
 \av{\delta a^{\dagger}_{\mathrm{in}}(\omega) \delta a^{\dagger}_{\mathrm{in}}(\omega')} \\&=-2\pi M\delta(\omega+\omega'),
\end{split}
\end{align}
one finds the quadrature noise correlations
\begin{align}\label{eq:28}
\begin{split}
    \av{\delta q_{\mathrm{in}}(\omega) \delta q_{\mathrm{in}}(\omega')} =&\bigg(N-M+\frac{1}{2}\bigg)2\pi \delta(\omega+\omega'),\\
    \av{\delta p_{\mathrm{in}}(\omega) \delta p_{\mathrm{in}}(\omega')}=&\bigg(N+M+\frac{1}{2}\bigg)2\pi \delta(\omega+\omega'),\\
 \av{\delta q_{\mathrm{in}}(\omega) \delta p_{\mathrm{in}}(\omega')} =&  -\av{\delta p_{\mathrm{in}}(\omega) \delta q_{\mathrm{in}}(\omega')} \\&=\frac{i}{2}2\pi\delta(\omega+\omega').
 \end{split}
\end{align}

\begin{figure}[tbp]
\centering
\includegraphics[width=\linewidth]{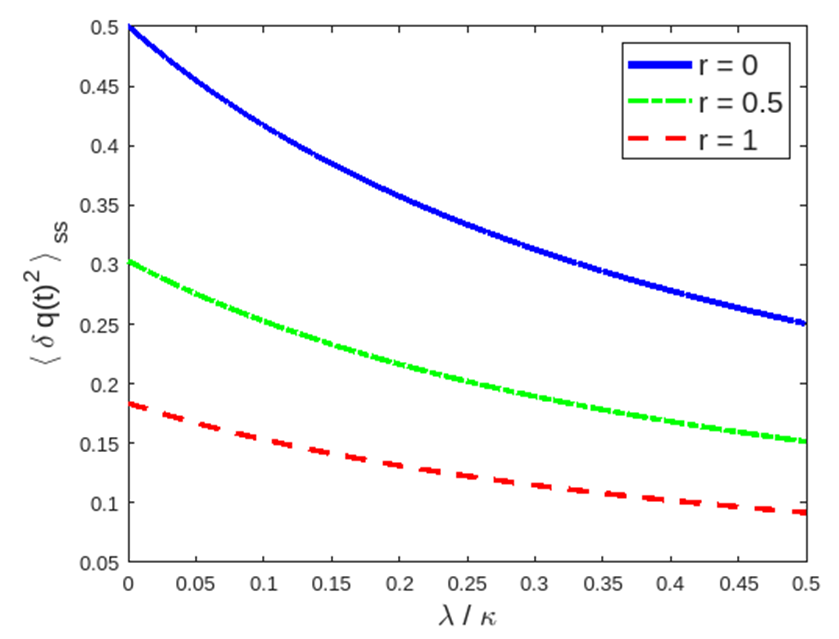}
\caption{Steady-state intracavity amplitude-quadrature variance $\av{\delta q(t)^2}_{\mathrm{ss}}$ as a function of $\lambda/\kappa$ for several reservoir squeeze parameters $r$, with the non-Hermitian contributions set to zero.}
\label{fig:1}
\end{figure}
Thus, substituting Eqs.~\eqref{eq:23} and \eqref{eq:28} into Eqs.~\eqref{eq:25} and \eqref{eq:26}, we obtain the spectra of fluctuations in the quadratures of the amplitude and phase of the cavity field as,
\begin{align}
    S_{q} (\omega) =&A_1(\omega)A_1(-\omega)\bigg(N-M+\frac{1}{2}\bigg)\nonumber\\&+B_1(\omega)B_1(-\omega)\bigg(N+M+\frac{1}{2}\bigg)\nonumber\\
    =& \frac{\frac{1}{2}\kappa[J^2_{22}+\omega^2]e^{-2r}+\frac{1}{2}\kappa J^2_{12}e^{2r}}{(J^2_{11}+\omega^2)(J^2_{22}+\omega^2)-2(J_{11}J_{22}-\omega^2)J_{12}J_{21}} ,\label{eq:29}\\
     S_{p} (\omega) =&A_2(\omega)A_2(-\omega)\bigg(N-M+\frac{1}{2}\bigg) \nonumber\\&+B_2(\omega)B_2(-\omega)\bigg(N+M+\frac{1}{2}\bigg),\nonumber\\
 =&\frac{\frac{1}{2}\kappa[J^2_{11}+\omega^2]e^{2r}+\frac{1}{2}\kappa J^2_{21}e^{-2r}}{(J^2_{11}+\omega^2)(J^2_{22}+\omega^2)-2(J_{11}J_{22}-\omega^2)J_{12}J_{21}}.\label{eq:30}
\end{align}

Therefore, the mean square fluctuations in the quadratures of the cavity field at steady state, $\av{\delta q(t)^2}_{ss}$ and $\av{\delta p(t)^2}_{ss}$ are given by
\begin{align}\label{eq:31}
   \av{\delta q(t)^2}_{ss}  =&\frac{1}{2\pi}\int_{-\infty}^{+\infty}d\omega S_{q} (\omega)\nonumber\\
   =&\frac{\kappa e^{-r}}{4\pi}\int_{-\infty}^{+\infty} \frac{\omega^2 d\omega}{\omega^4+b\omega^2 +c} \nonumber\\&+\frac{\kappa (e^{-r} J^2_{22}+ e^{r}J^2_{12})}{4\pi}\int_{-\infty}^{+\infty}\frac{d\omega}{\omega^4+b\omega^2 +c} ,\nonumber\\
   =&\frac{1}{4}\kappa e^{-r}\bigg(\frac{\sqrt{g}-\sqrt{f}}{h}\bigg) \nonumber\\&+\frac{\kappa (e^{-r} J^2_{22}+ e^{r}J^2_{12})}{4} \bigg(\frac{g\sqrt{f}-f\sqrt{g}}{hfg}\bigg)
\end{align}
\begin{align}\label{eq:32}
 \av{\delta p(t)^2}_{ss} =&\frac{1}{2\pi}\int_{-\infty}^{+\infty}d\omega S_{p}(\omega),\nonumber\\
   =&\frac{\kappa e^{r}}{4\pi}\int_{-\infty}^{+\infty} \frac{\omega^2 d\omega}{\omega^4+b\omega^2 +c}\nonumber\\&+\frac{\kappa (e^{r} J^2_{11}+ e^{-r}J^2_{21})}{4\pi}\int_{-\infty}^{+\infty}\frac{d\omega}{\omega^4+b\omega^2 +c} ,\nonumber\\
    =&\frac{1}{4}\kappa e^{r}\bigg(\frac{\sqrt{g}-\sqrt{f}}{h}\bigg) \nonumber\\&+\frac{\kappa (e^{r} J^2_{11}+ e^{-r}J^2_{21})}{4} \bigg(\frac{g\sqrt{f}-f\sqrt{g}}{hfg}\bigg),
\end{align}
where $b=(J^2_{11}+J^2_{22}+2J_{12}J_{21}), c =J_{11}J_{22}(J_{11}J_{22}-2J_{12}J_{21}), h=\sqrt{b^2-4c}, g=\frac{1}{2}(b+h)$, and $f=\frac{1}{2}(b-h)$.

As a consistency check, we now set $\gamma=\eta=\chi=0$. In this Hermitian limit, the following spectra and variances reduce to the previously published results for a coherently driven DOPO coupled to a squeezed-vacuum reservoir \cite{TEKLU2006}. We retain them here to make explicit how the present generalized formulae connect to the standard case.
Equation~(\ref{eq:29}) and Eq.~\eqref{eq:30} for the squeezing spectra reduced to
\begin{align}\label{eq:33}
\begin{split}
    S_{q} (\omega)  =& \frac{\frac{1}{2}\kappa e^{-2r}}{(\frac{\kappa}{2}+\lambda)^2+\omega^2} ,
    \\
 S_{p} (\omega) =&\frac{\frac{1}{2}\kappa e^{2r}}{(\frac{\kappa}{2}-\lambda)^2+\omega^2},
\end{split}
\end{align}
and the mean square quadrature fluctuations Eq.~\eqref{eq:31} and Eq.~\eqref{eq:32} become
\begin{align}\label{eq:34}
\begin{split}
   \av{\delta q(t)^2}_{ss}  =& \frac{\frac{1}{2}\kappa e^{-2r}}{\kappa+2\lambda},    \\
   \av{\delta p(t)^2}_{ss} =& \frac{\frac{1}{2}\kappa e^{2r}}{\kappa-2\lambda},
\end{split}
\end{align}
Here, we also observe that, at steady state, the driving coherent light does not affect the quadrature variances. Moreover, in the absence of a squeezed vacuum reservoir and a parametric oscillator, the quadrature squeezing has the same value as the vacuum noise level, $\av{\delta q(t)^2}=\av{\delta p(t)^2}=0.5$. The cavity field exhibits amplitude-quadrature squeezing.
In the present below-threshold DOPO, the ``threshold'' is the point at which the linear damping of one quadrature vanishes and a drift-matrix eigenvalue reaches zero. In the Hermitian limit this gives $\lambda_{\rm th}=\kappa/2$, or equivalently $\kappa=2\lambda$. Below threshold, $\lambda<\lambda_{\rm th}$, the linearized steady state is stable; at threshold, the ideal linearized model predicts critical behavior.
At the threshold condition $\kappa=2\lambda$, it will have,
\begin{align}\label{eq:35}
\begin{split}
   \av{\delta q(t)^2}_{ss}  =& \frac{1}{4} e^{-2r},    \\
\av{\delta p(t)^2}_{ss} \to & \infty,
\end{split}
\end{align}

Expression~\eqref{eq:35} shows that, at steady state and at threshold, the variance of the amplitude quadrature equals the product of the variance obtained when the DOPO is coupled to an ordinary vacuum reservoir and the variance associated with the squeezed vacuum reservoir. Thus, the squeezed reservoir enhances the cavity mode's degree of squeezing, while the coherent drive does not affect the quadrature variances.
The divergence of the phase-quadrature variance in Eq.~\eqref{eq:35} should not be read as a claim of experimentally attainable infinite noise or perfect squeezing. It is the critical singularity of the ideal below-threshold, linearized model. In real devices, it is regularized by pump depletion, nonlinear saturation above threshold, finite detection bandwidth, intracavity loss, finite reservoir bandwidth, and technical noise. Operationally, it marks the point at which the linearized steady-state approximation ceases to be sufficient.

\begin{figure}[tbp]
\centering
\includegraphics[width=\linewidth]{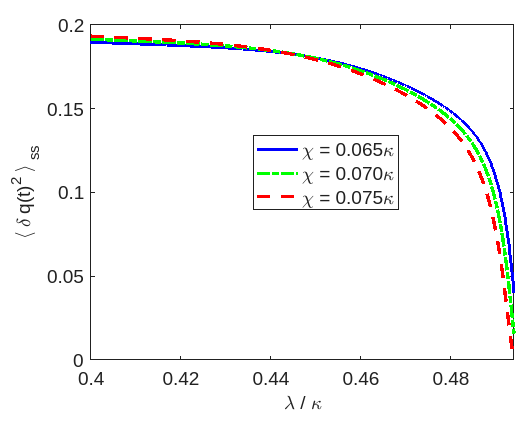}
\caption{Steady-state intracavity amplitude-quadrature variance $\av{\delta q(t)^2}_{\mathrm{ss}}$ as a function of $\lambda/\kappa$ for several values of $\chi$, with $\eta=0.03\kappa$ and $r=0.5$.}
\label{fig:2}
\end{figure}

Figure~\ref{fig:1} illustrates that the quadrature variance decreases with both the nonlinear gain \(\lambda\) and the squeeze parameter \(r\). For \(\lambda =0\), squeezing is already present due to the squeezed reservoir. In the absence of reservoir engineering and nonlinear intracavity interaction, the squeezing remains at the vacuum level. Optimal squeezing occurs near the threshold condition for each value of $r$.

\begin{figure}[tbp]
\centering
\includegraphics[width=\linewidth]{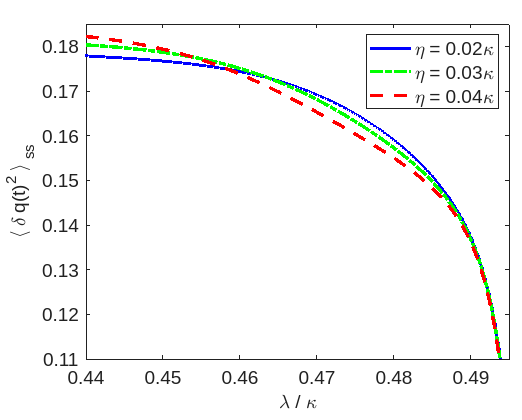}
\caption{Steady-state intracavity amplitude-quadrature variance $\av{\delta q(t)^2}_{\mathrm{ss}}$ as a function of $\lambda/\kappa$ for several values of $\eta$, with $\chi=0.05\kappa$ and $r=0.5$.}
\label{fig:3}
\end{figure}

Figures~\ref{fig:2} and~\ref{fig:3} show the impact of non-Hermitian parameters. In Fig.~\ref{fig:2}, the loss term associated with the parametric pumping $\chi$ has a positive effect on the quadrature variance beyond a certain nonlinear gain, especially near threshold. In Fig.~\ref{fig:3}, increasing the loss parameter $\eta$ associated with the coherent driving initially reduces squeezing up to a certain value of $\lambda$. However, near threshold, it can have a constructive effect on the amplitude-quadrature variance over a specific range of $\lambda$.

The reduced sensitivity to the NH parameters as $\lambda$ approaches threshold has a simple physical origin. The amplitude quadrature is the squeezed quadrature of the DOPO, and its effective damping contains the dominant Hermitian contribution $\kappa/2+\lambda$. Near threshold, this scale is of order $\kappa$, whereas the NH corrections used here are calibrated shifts of the drift coefficients that are smaller. Consequently, close to threshold, the amplitude-quadrature variance is controlled mainly by the squeezed reservoir and by the large Hermitian amplitude-quadrature damping, while $\chi$ and $\eta$ act mostly as finite renormalizations of the drift matrix.

\section{Squeezing spectrum}\label{sec:4}

\begin{figure}[tbp]
\centering
\includegraphics[width=\linewidth]{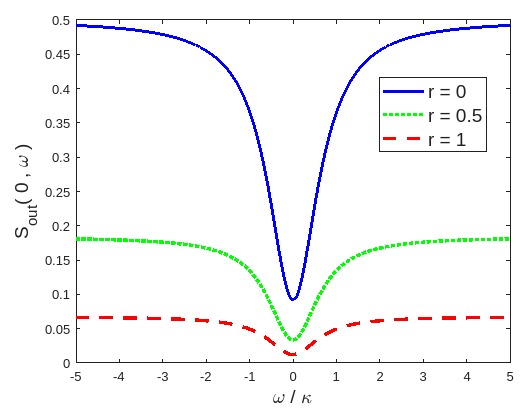}
\caption{Output amplitude-quadrature squeezing spectrum $S_{\mathrm{out}}(0,\omega)$ as a function of $\omega/\kappa$ for several squeeze parameters $r$, with $\lambda=0.2\kappa$ and the non-Hermitian contributions set to zero.}
\label{fig:4}
\end{figure}

We now analyze the squeezing of the cavity mode as inferred from the output field. The fluctuation $\delta \hat a (\omega)$ of the cavity field follows from Eq.~\eqref{eq:19}. Using the standard input--output relation \cite{walls2008input},
$\hat a_{\mathrm{out}}(t) = \sqrt{\kappa} \hat a (t)-\hat a_{\mathrm{in}}(t)$, one obtains the fluctuation  $\delta\hat a_{\mathrm{out}}(\omega)$ of the output cavity field. Then, we define the quadrature fluctuation of the output cavity field as
\begin{align}\label{eq:36}
\delta\hat R_{\mathrm{out}}(\omega)
&=\frac{1}{\sqrt{2}}\left[
\delta\hat a_{\mathrm{out}}(\omega)e^{-i\theta}
+\delta\hat a_{\mathrm{out}}^{\dagger}(\omega)e^{i\theta}
\right]\nonumber\\
&=\delta\hat q_{\mathrm{out}}(\omega)\cos\theta
+\delta\hat p_{\mathrm{out}}(\omega)\sin\theta.
\end{align}
Here $\theta$ is the phase of the local oscillator, with
$\delta\hat q_{\mathrm{out}}(\omega)=
[\delta\hat a_{\mathrm{out}}^{\dagger}(\omega)+
\delta\hat a_{\mathrm{out}}(\omega)]/\sqrt{2}$ and
$\delta\hat p_{\mathrm{out}}(\omega)=
i[\delta\hat a_{\mathrm{out}}^{\dagger}(\omega)-
\delta\hat a_{\mathrm{out}}(\omega)]/\sqrt{2}$.
For $\theta=0$, $\delta\hat R_{\mathrm{out}}=\delta\hat q_{\mathrm{out}}$ is the amplitude fluctuation; for $\theta=\pi/2$, $\delta\hat R_{\mathrm{out}}=\delta\hat p_{\mathrm{out}}$ is the phase fluctuation. Using the input--output relation together with Eq.~\eqref{eq:22}, we obtain
\begin{align}\label{eq:37}
\mathbf{u}_{\mathrm{out}}(\omega)
=\left[\sqrt{\kappa}\left(-i\omega\mathbf{I}-\mathbf{J}\right)^{-1}
-\frac{1}{\sqrt{\kappa}}\mathbf{I}\right]\mathbf{n}(\omega),
\end{align}
where $\mathbf{u}_{\mathrm{out}}(\omega)=
[\delta\hat q_{\mathrm{out}}(\omega),
\delta\hat p_{\mathrm{out}}(\omega)]^{\mathsf T}$.
The output quadrature can then be written as
\begin{align}\label{eq:38}
\delta\hat R_{\mathrm{out}}(\omega)
=R_q(\omega)\hat q_{\mathrm{in}}(\omega)
+R_p(\omega)\hat p_{\mathrm{in}}(\omega),
\end{align}
where, using the definition of $D(\omega)$ below Eq.~\eqref{eq:24},
\begin{align*}
R_q(\omega)
&=\frac{\kappa J_{21}}{D(\omega)}\sin\theta\nonumber\\[-0.25em]
&\quad-\left[1+\frac{\kappa(J_{22}+i\omega)}{D(\omega)}\right]\cos\theta,\\
R_p(\omega)
&=\frac{\kappa J_{12}}{D(\omega)}\cos\theta\nonumber\\[-0.25em]
&\quad-\left[1+\frac{\kappa(J_{11}+i\omega)}{D(\omega)}\right]\sin\theta.
\end{align*}
For compactness, let $X_{\theta}(\omega)\equiv
\delta\hat R_{\mathrm{out}}(\omega)$. The output-quadrature spectrum is defined by
\begin{align}\label{eq:39}
S_{\mathrm{out}}(\theta,\omega)
&=\frac{1}{4\pi}\int_{-\infty}^{\infty}d\omega'\,
 e^{i(\omega+\omega')t}\nonumber\\
&\quad\times\Bigl[
\av{X_{\theta}(\omega)X_{\theta}(\omega')}\nonumber\\
&\qquad+\av{X_{\theta}(\omega')X_{\theta}(\omega)}\Bigr].
\end{align}
Substituting Eqs.~\eqref{eq:28} and \eqref{eq:38} into Eq.~\eqref{eq:39} gives
\begin{align}\label{eq:40}
S_{\mathrm{out}}(\theta,\omega)
&=R_q(\omega)R_q(-\omega)\left(N-M+\frac{1}{2}\right)\nonumber\\
&\quad+R_p(\omega)R_p(-\omega)\left(N+M+\frac{1}{2}\right)\nonumber\\
&=\frac{1}{2}R_q(\omega)R_q(-\omega)e^{-2r}
+\frac{1}{2}R_p(\omega)R_p(-\omega)e^{2r}.
\end{align}
The output field is in a squeezed state when $S_{\mathrm{out}} (\theta,\omega)<1/2$, i.e., below the vacuum level.
Furthermore, in the absence of non-Hermitian Hamiltonian contributions, $S_{\mathrm{out}}(\theta,\omega)$ in Eq.~\eqref{eq:40} reduces to the standard Hermitian expression reported in Ref.~\cite{TEKLU2006},
\begin{align}\label{eq:41}
S_{\mathrm{out}}(\theta,\omega)=& \bigg(\frac{1}{2}-\frac{4\kappa\lambda }{(\kappa+2\lambda)^2+4\omega^2}\bigg)e^{-2r}\cos^2{\theta}\nonumber\\&+\bigg(\frac{1}{2}+\frac{4\kappa\lambda }{(\kappa-2\lambda)^2+4\omega^2}\bigg)e^{2r}\sin^2{\theta},
\end{align}
from which the amplitude and phase spectra follow:
\begin{align}\label{eq:42}
S_{\mathrm{out}}(0,\omega)=& \bigg(\frac{1}{2}-\frac{4\kappa\lambda }{(\kappa+2\lambda)^2+4\omega^2}\bigg)e^{-2r},
\end{align}
\begin{align}\label{eq:43}
S_{\mathrm{out}}(\pi/2,\omega)=& \bigg(\frac{1}{2}+\frac{4\kappa\lambda}{(\kappa-2\lambda)^2+4\omega^2}\bigg)e^{2r}.
  \end{align}
Since the stability of the system in the absence of non-Hermitian Hamiltonian parts is $\lambda<\kappa/2$, and at threshold condition where $(\kappa=2\lambda)$ the output squeezing spectra $S_{\mathrm{out}}(0,\omega)$ and $S_{\mathrm{out}}(\pi/2,\omega)$ become
\begin{align}\label{eq:44}
S_{\mathrm{out}}(0,\omega)=& \frac{1}{2}\bigg(\frac{\omega^2 }{\kappa^2+\omega^2}\bigg)e^{-2r},
   \end{align}
\begin{align}\label{eq:45}
S_{\mathrm{out}}(\pi/2,\omega)=& \frac{1}{2}\bigg(\frac{\kappa^2+\omega^2 }{\omega^2}\bigg)e^{2r}.
   \end{align}

Thus, the squeezed reservoir enhances the output squeezing: for $\omega=0$, the amplitude quadrature noise can be totally suppressed, while the phase quadrature diverges. As in the intracavity case, the coherent driving does not affect the squeezing spectrum.

\begin{figure}[tbp]
\centering
\includegraphics[width=\linewidth]{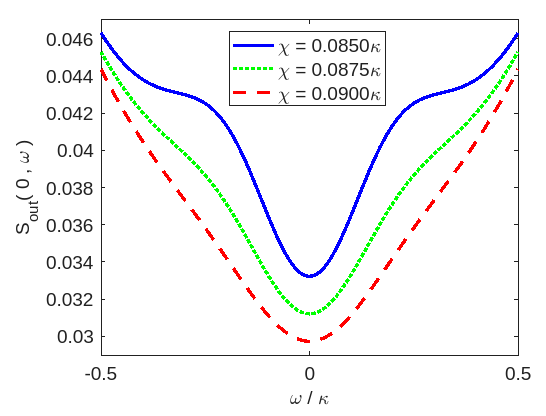}
\caption{Output amplitude-quadrature squeezing spectrum $S_{\mathrm{out}}(0,\omega)$ as a function of $\omega/\kappa$ for several values of $\chi$, with $\eta=0.035\kappa$, $\varepsilon=0.3\kappa$, $r=0.5$, and $\lambda=0.48\kappa$.}
\label{fig:5}
\end{figure}

\begin{figure}[tbp]
\centering
\includegraphics[width=\linewidth]{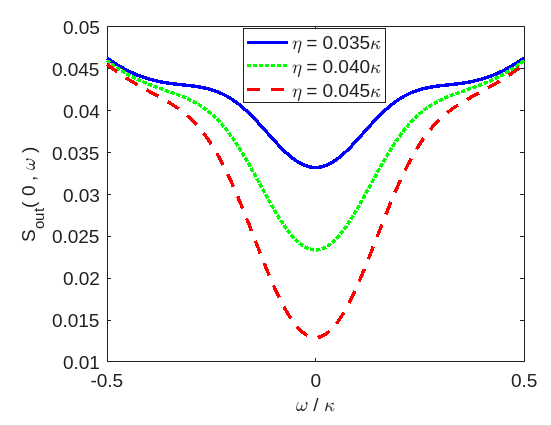}
\caption{Output amplitude-quadrature squeezing spectrum $S_{\mathrm{out}}(0,\omega)$ as a function of $\omega/\kappa$ for several values of $\eta$, with $\chi=0.085\kappa$, $\varepsilon=0.3\kappa$, $r=0.5$, and $\lambda=0.48\kappa$.}
\label{fig:6}
\end{figure}
With or without the NH Hamiltonian, output squeezing appears as suppression of the amplitude-quadrature spectrum below the vacuum level. Figure~\ref{fig:4} shows that the strongest suppression occurs at resonance, $\omega=0$, for all values of $r$. Increasing the squeezed-reservoir strength lowers the amplitude-quadrature spectrum and broadens the frequency region over which the output remains squeezed.

Figures~\ref{fig:5} and~\ref{fig:6} show that, for the stable parameter sets considered here, the NH drift terms can further reduce the resonant output noise. The enhancement is most visible near $\omega=0$ because the resonant output field has the longest effective interaction time with the parametrically squeezed intracavity mode. Moving away from resonance, the finite detuning term in the denominator of the response functions competes with the drift-matrix shifts, so the relative influence of $\chi$ and $\eta$ decreases.

\section{Power spectrum} \label{sec:5}

\begin{figure}[tbp]
\centering
\includegraphics[width=\linewidth]{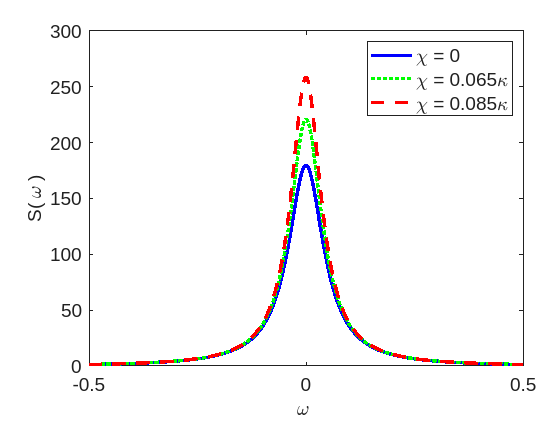}
\caption{Cavity-mode power spectrum $S(\omega)$ as a function of $\omega$ for several values of $\chi$, with $\eta=0$, $\lambda=0.45\kappa$, $\varepsilon=0$, and $r=0$.}
\label{fig:7}
\end{figure}
We now calculate the power spectrum of the cavity mode. The power spectrum of a single-mode light is expressible as the Fourier transform of the two-time correlation functions,
 $\big<\delta \hat a^{\dagger}(t+\tau) \delta \hat a (t)\big>$:
\begin{align}\label{eq:46}
S(\omega)
&=\int_{-\infty}^{\infty}d\tau\,e^{-i\omega\tau}
\av{\delta\hat a^{\dagger}(t+\tau)\delta\hat a(t)}
\equiv C_{\hat a^{\dagger}\hat a}(\omega).
\end{align}
Using $S(\omega)\equiv C_{\hat a^{\dagger}\hat a}(\omega)$ from Eq.~\eqref{eq:46}, the corresponding symmetrized frequency-domain correlation is
\begin{align}\label{eq:47}
S(\omega)
&=\frac{1}{4\pi}\int_{-\infty}^{\infty}d\omega'\,
 e^{i(\omega+\omega')t}\nonumber\\
&\quad\times\Bigl[
\av{\delta\hat a^{\dagger}(\omega)\delta\hat a(\omega')}\nonumber\\
&\qquad+\av{\delta\hat a^{\dagger}(\omega')\delta\hat a(\omega)}\Bigr],\nonumber\\
2\pi S(\omega)\delta(\omega+\omega')
&=\frac{1}{2}\Bigl[
\av{\delta\hat a^{\dagger}(\omega)\delta\hat a(\omega')}\nonumber\\
&\qquad+\av{\delta\hat a^{\dagger}(\omega')\delta\hat a(\omega)}\Bigr].
\end{align}
Using $\delta\hat a(\omega)=
[\delta\hat q(\omega)+i\delta\hat p(\omega)]/\sqrt{2}$ and Eq.~\eqref{eq:28}, we obtain
\begin{align}\label{eq:48}
\delta\hat a(\omega)
=\frac{1}{\sqrt{2}}\left[
 f_1(\omega)\hat q_{\mathrm{in}}(\omega)
+f_2(\omega)\hat p_{\mathrm{in}}(\omega)
\right].
\end{align}
Using $\delta\hat a^{\dagger}(\omega)=
[\delta\hat a(-\omega)]^{\dagger}$ and
$\int_{-\infty}^{\infty}g^*(-\tau)e^{-i\omega\tau}d\tau=g^*(-\omega)$ gives
\begin{align}\label{eq:49}
\delta\hat a^{\dagger}(\omega)
=\frac{1}{\sqrt{2}}\left[
 f_1^*(-\omega)\hat q_{\mathrm{in}}(\omega)
+f_2^*(-\omega)\hat p_{\mathrm{in}}(\omega)
\right].
\end{align}
Here $f_1(\omega)=A_1(\omega)+iA_2(\omega)$ and $f_2(\omega)=B_1(\omega)+iB_2(\omega)$. Substituting Eqs.~\eqref{eq:48} and \eqref{eq:49} into Eq.~\eqref{eq:47}, and using Eq.~\eqref{eq:28}, gives the power spectrum
\begin{align}\label{eq:50}
     S(\omega) =&\frac{1}{4}\bigg[\big[\abs{f_1(-\omega)}^2+\abs{f_1(\omega)}^2-1\big]e^{-2r}\nonumber\\&+ \big[\abs{f_2(-\omega)}^2+\abs{f_2(\omega)}^2+1\big]e^{2r} \nonumber\\ & +
     if^{*}_1(-\omega)f_2(-\omega)-if^{*}_2(-\omega)f_1(-\omega)\nonumber\\&+if^{*}_1(\omega)f_2(\omega)-if^{*}_2(\omega)f_1(\omega)\bigg].
   \end{align}

We will evaluate the power spectrum of the cavity mode in the absence of the driving radiation ($\varepsilon=0$ and hence $\eta=0$), and we also note that for $r=0$, Eq.~\eqref{eq:50} reduces to the power spectrum of a degenerate parametric oscillator coupled to an ordinary vacuum reservoir,
\begin{align}\label{eq:51}
     S(\omega) =&\frac{1}{4}\bigg[\abs{f_1(-\omega)}^2+\abs{f_1(\omega)}^2+ \abs{f_2(-\omega)}^2+\abs{f_2(\omega)}^2\nonumber \\ & +
     if^{*}_1(-\omega)f_2(-\omega)-if^{*}_2(-\omega)f_1(-\omega)\nonumber\\&+if^{*}_1(\omega)f_2(\omega)-if^{*}_2(\omega)f_1(\omega)\bigg].
   \end{align}
In the absence of non-Hermitian parameters, $f_1(\omega)=A_1(\omega)=-\frac{\sqrt{\kappa}}{D(\omega)}[-\frac{\kappa}{2}+\lambda+i\omega]$, and $f_2(\omega)=i B_2(\omega)=-\frac{i\sqrt{\kappa}}{D(\omega)}[-\frac{\kappa}{2}-\lambda+i\omega]$, we obtain the previously known Hermitian DOPO expression~\cite{TEKLU2006}:
\begin{align}\label{eq:52}
     S(\omega) =&\frac{1}{4}\bigg[\abs{f_1(-\omega)}^2+\abs{f_1(\omega)}^2+ \abs{f_2(-\omega)}^2+\abs{f_2(\omega)}^2 \bigg]\nonumber\\&
     =\frac{\kappa}{2}\bigg[\frac{1}{(\frac{\kappa}{2}+\lambda)^2+\omega^2}+\frac{1}{(\frac{\kappa}{2}-\lambda)^2+\omega^2}\bigg].
   \end{align}
Figure~\ref{fig:7} displays the power spectrum of the cavity mode in the absence and in the presence of non-Hermitian contributions. In both cases, the maximum of the spectrum occurs at $\omega=0$. Overall, the intracavity parametric pumping loss has a positive effect on the power spectrum near $\omega=0$, thereby enhancing its peak.
This behavior follows from the Lorentzian-like response of the linearized cavity: at exact resonance, the field samples the modified drift rates for the longest time. Hence, a small NH change in the effective linewidth produces a visible peak-height change. For nonzero detuning, the term $\omega^2$ in the response denominator increasingly dominates over the small drift-rate shifts due to $\chi$ and $\eta$. Therefore, spectra with different NH parameters tend to converge away from the resonance, and the NH contribution becomes less pronounced.

\section{Conclusion}\label{sec:6}
We have investigated environment-assisted parametric amplification with non-Hermitian Hamiltonian effects on the intracavity quadrature squeezing, the output squeezing spectrum, and the power spectrum of a degenerate parametric oscillator driven by coherent light and coupled to a squeezed vacuum reservoir. The various non-Hermitian parameters have distinct impacts on the intracavity and output properties of the cavity mode. In particular, non-Hermitian contributions can open parameter windows in which the amplitude-quadrature variance falls below the standard quantum limit more strongly than in the Hermitian case, and they can substantially modify the output squeezing and power spectra near optimal operating points.

These features suggest that non-Hermitian control, combined with reservoir engineering and coherent driving, offers a promising route to tailoring nonclassical light in realistic, lossy platforms. The resulting irregularities and enhancements in squeezing under near-threshold conditions may be exploited in quantum metrology applications, for example, to improve the sensitivity of precision measurements.
At the same time, the NH parameters should be regarded as calibrated effective drift coefficients, and a physical implementation must include the accompanying quantum-noise channels and satisfy the stability conditions stated above. The present results, therefore, provide parameter windows and diagnostic trends for experiments, rather than a claim that arbitrary NH coefficients can be added without constraints.

\begin{acknowledgments}
This work was supported by Khalifa University of Science and Technology through project 8474000739 (RIG-2024-033). The work was facilitated by the support and resources provided by Adama Science and Technology University. The research of K.G.Z. was supported by the Department of Higher Education and Training of South Africa and in part by the National Research Foundation of South Africa (Grants No. 95965 and No. 132202).
\end{acknowledgments}

\appendix

\section{Density matrix formalism for non-Hermitian Hamiltonians}\label{appendix}

In this appendix we briefly recall the density-matrix formulation of non-Hermitian quantum dynamics. We focus on the most general types of dynamics that involve both pure and mixed states. Upon introducing a non-normalized reduced density matrix \(\hat \rho\), such dynamics can be described by a master equation of the form \cite{fbook}
\begin{equation}
\partial_t {\hat \rho} =-\frac{i}{\hbar}\left[\hat H, \hat \rho\right]-
\frac{1}{\hbar}
\left\{\hat{\Gamma},\hat \rho \right\}
+
\hat{\cal D} ({\hat \rho}, \hat A ),
\label{e1}
\end{equation}
where \([ , ]\) and \(\{ , \}\) denote the commutator and anti-commutator, respectively, and \(\hat{\cal D} ({\hat \rho},  \hat A )\) is the dissipator. The dissipator is linear in \(\hat \rho (t)\), quadratic in the operators \(\hat A\), and obeys the quantum dynamical semigroup form \cite{bpbook}. It must be traceless in order for the trace of \(\hat \rho\) to be conserved under the action of \(\hat{\cal D}\).

In the context of open quantum systems, Eq.~\eqref{e1} effectively describes a subsystem with Hamiltonian $\hat{H}$ coupled to an environment represented
by $\hat{\Gamma}$. Taking the trace of both sides of Eq.~\eqref{e1} yields
\begin{equation}
\partial_t{\rm Tr}\,\hat \rho
=-\frac{2}{\hbar}{\rm Tr}(\hat{\Gamma} \,\hat \rho ),
\label{eq:dotTrOmega}
\end{equation}
which shows that \({\rm Tr}\,\hat \rho\) is not generally conserved; the non-Hermitian dynamics does not preserve the probability measure. As suggested in Ref.~\cite{sz13}, one introduces a normalized density operator
\begin{equation}
\hat \rho'
 =\hat \rho /{\rm Tr}\,\hat \rho,
\label{e:rho}
\end{equation}
which can be used to compute expectation values of physical observables:
\(\av{\hat A}' = \text{Tr} (\hat A \hat \rho')
= \text{Tr} (\hat A \hat \rho)/\text{Tr} \hat \rho
=\av{\hat A}/\text{Tr} \hat \rho\).
Here \(\hat A\) is an operator associated with an observable. As a result of Eqs.~\eqref{e1} and \eqref{e:rho}, the normalized density operator obeys
\begin{equation}
    \partial_t {\hat \rho'} =
-\frac{i}{\hbar}\left[\hat H, \hat \rho'\right]
-\frac{1}{\hbar}\left\{\hat{\Gamma},\hat \rho' \right\}
+\frac{2}{\hbar}\hat \rho' \, {\rm Tr} (\hat{\Gamma}\hat \rho')
+\hat{\cal D} ({\hat \rho'},  \hat A ),
\label{eq:dotrho}
\end{equation}
which is nonlinear and nonlocal in $\hat \rho'$ \cite{sz14,z24uni}.
The operator $\hat \rho'$ is bounded and allows one to maintain a probabilistic interpretation of expectation values under non-Hermitian dynamics. At the same time, the gain or loss of probability associated with the coupling to sinks or sources is naturally described by the non-normalized density operator $\hat \rho$. Depending on the physical context and boundary conditions, one can use either $\hat \rho$ or $\hat \rho'$; see, e.g., Refs.~\cite{z16,z24jmo} for explicit examples.

\bibliographystyle{apsrev4-2}
\bibliography{apssamp}

\end{document}